\documentclass[proof]{WileyASNA-v1}

\articletype{review}%

\received{}
\revised{}
\accepted{}

\begin{document}

\title{{Optical and UV properties of a radio-loud and a radio-quiet Population A quasar at high redshift}\protect\thanks{Invited talk presented at the 13$^{\rm{th}}$ Serbian Conference on spectral line shapes in astrophysics (SCSLSA)}}

\author[1]{A. Deconto-Machado*}

\author[1]{A. del Olmo}

\author[2]{P. Marziani}

\author[1]{J. Perea}

\author[3]{G. M. Stirpe}

\authormark{A. DECONTO-MACHADO \textsc{et al}}

\address[1]{\orgname{Instituto de Astrofísica de Andalucía (IAA-CSIC)}, \orgaddress{\state{Granada}, \country{Spain}}}

\address[2]{\orgname{INAF, Osservatorio Astronomico di Padova}, \orgaddress{\state{Padova}, \country{Italy}}}

\address[3]{\orgname{INAF, Osservatorio di Astrofisica e Scienza dello Spazio}, \orgaddress{\state{Bologna}, \country{Italy}}}

\corres{Alice Deconto Machado, Instituto de Astrofísica de Andalucía (IAA-CSIC), Glorieta de la Astronomía, S/N, 18008 Granada, Spain. \email{adeconto@iaa.es}}

\abstract{Different properties of quasars may be observed and analysed through the many ranges of the electromagnetic spectrum. Pioneering studies showed that an ``H-R diagram'' for quasars was needed to organize these data, and that more than two dimensions were necessary: a four dimensional Eigenvector (4DE1) parameter space was proposed. The 4DE1 makes use of independent observational properties obtained from the optical and UV emission lines, as well as from the soft-X rays. The 4DE1 ``optical plane'', also known as the quasar Main Sequence (MS), identifies different spectral types in order to describe a consistent picture of QSOs. In this work we present a spectroscopic analysis focused on the comparison between two sources, one radio-loud (PKS2000-330, $z=3.7899$) and one radio-quiet (Q1410+096, $z=3.3240$), both showing Population A quasar spectral properties. Optical spectra were observed in the infrared with VLT/ESO, and the additional measures in UV were obtained through the fitting of archive spectra. The analysis was performed through a non-linear multi-component decomposition of the emission line profiles. Results are shown in order to highlight the effects of the radio-loudness on their emission line properties. The two quasars share similar optical spectroscopic properties and are very close on the MS classification while presenting significant differences on the UV data. Both sources show significant blueshifts in the UV lines but important differences in their UV general behaviour. While the radio-quiet source Q1410+096 shows a typical Pop A UV spectrum with similar intensities and shapes on both C\textsc{IV}$\lambda$1549 and Si\textsc{IV}$\lambda$1392, the UV spectrum of the strong radio-loud PKS2000-330 closely resembles the one of population B of quasars.} 






\maketitle

\begin{center}
\begin{table*}[h!]
\centering
\caption{Main properties of PKS2000-330 and Q1410+096.\label{tab2}}%
\tabcolsep=0pt%
\begin{tabular*}{40pc}{@{\extracolsep\fill}lccccr@{\extracolsep\fill}}
\toprule
\textbf{Source} & \textbf{RA (J2000)}  & \textbf{DEC (J2000)} & \textbf{$z$} & \textbf{$M_{\rm{i}}$} & \textbf{$m_{\rm{H}}$}\\
(1) & (2) & (3) & (4) & (5) & (6)\\
\midrule
\textbf{Q1410+096} & 14 13 21.05  & +09 22 04.8 & 3.3240 & -29.44 & 15.62 \\
\textbf{PKS2000-330} & 20 02 24.00  & -32 51 47.0 & 3.7899 & -30.99 & 15.32\\

\bottomrule
\label{tab:properties}
\end{tabular*}
\end{table*}
\end{center}

\section{Introduction}\label{sec1}
\par In order to organise the spectroscopic diversity observed in low-redshift quasars, \cite{Sulentic_2000} introduced  the fourth-dimensional Eigenvector 1 (4DE1), a correlation space that considers several key observational measures (including optical, UV, and X-ray) as well as physical parameters such as outflow prominence and accretion mode. One of the most important measures is the full width at half maximum (FWHM) of the H$\beta$ broad component, thought to be a measure of virialized motions in the accretion disc and thus crucial for black hole mass estimations. Also, another fundamental parameter is the ratio between the intensities of the Fe \textsc{II} blend at 4570\r{A} and H$\beta$ ($R_{\rm{Fe\textsc{II}}}=I(\textrm{Fe \textsc{II}}\lambda4570)/I($H$\beta)$), that is related to the Eddington ratio \citep{marziani_2001} and many of the optical and UV spectral line measures \citep{kovacevic_2010,Marziani_2010,shapovalova_2012}, and which could be used to estimate physical parameters of the Broad Line Region (BLR) as the ionisation state, the electron density or the column density \citep{panda_2020, ferland_2009}.

\par The optical plane of the 4DE1 (defined by the FWHM(H$\beta$) vs. R$_{Fe \rm{II}}$) is the so-called Main Sequence (MS) of quasars and the physical parameters related with accretion rate and outflowing gas seem to be changing along it \citep{Marziani_2018}. The MS gave rise to the concept of two populations of quasars, which present significant spectroscopic differences \citep{Zamfir_2010}. Along the Main Sequence, the physical properties vary from sources with low Fe \textsc{II} emission and high black hole mass ``disk-dominated'' QSOs (Population B), to the extreme Population A with strong Fe \textsc{II} emission, narrower Lorentzian line profiles, lower ionization spectra and evidence of strong outflows, ``wind-dominated'' quasars \citep{martinez_aldama_2018}. The optical plane of the 4DE1 is driven by the Eddington ratio convolved with orientation effects, and a critical Eddington ratio of $L/L_{\rm{Edd}}\approx 0.2$, associated with an accretion mode change, may play a key role in the observed MS \citep{Marziani_2019, sulentic_2017}.

\par Only 10\% of the known quasars are strong emitters in radio (radio-loud).  \cite{Zamfir_2010} analysed about 500 quasars observed with Sloan Digital Sky survey (SDSS) and found that radio-quiet sources are distributed equally in Pop. A and Pop. B. But this is not true for radio-loud sources. In the optical domain of the 4DE1, the radio-loud sources show a preference of having FWHM of H$\beta$ higher than 4000 km s$^{-1}$ and a R$_{\rm{Fe \textsc{II}}}$ lower than 0.5, which are the properties that characterize Pop. B quasars. However, recent results have found a relatively high fraction of intermediate radio-emitters in extreme Pop. A quasars \citep{del_olmo_2021, Ganci_2019}. These results may support a real dichotomy between radio-loud and radio-quiet QSOs. This work aims to contribute to that study by analysing the spectroscopic differences of two Population A quasars at high redshift, one radio-loud and one radio-quiet. Observational data and properties of the two quasars are described in Section 2. The performed optical and UV spectral analysis are presented in Sec. 3, while the main results are detailed in Sec. 4. Comparison of the spectroscopic properties of both sources and the main conclusions are presented in sections 5 and 6 respectively.


\section{Sample \& Observational Data}\label{sec2}
\begin{figure*}
    \centering
    \includegraphics[width=0.938\linewidth]{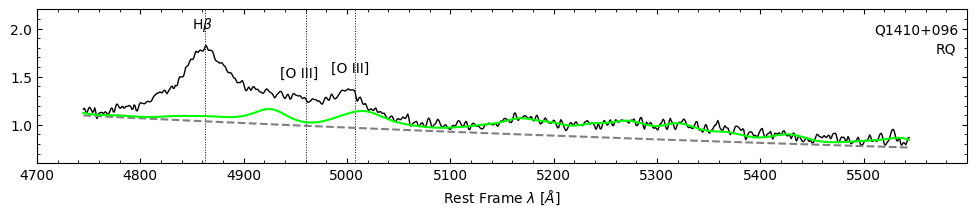}    \\
    \includegraphics[width=0.95\linewidth]{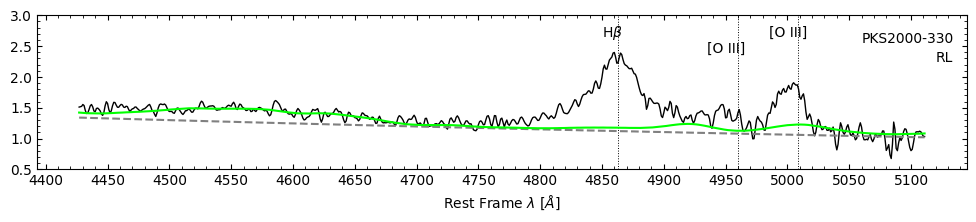}
    \caption{Optical spectra obtained with VLT/ISAAC for Q1410+036 (top panel) and PKS2000-330 (bottom panel). Grey dashed line indicates the powerlaw obtained on the \textsc{specfit} fitting to represent the continuum level. Green line shows the Fe\textsc{II} contribution. Vertical dotted lines indicate the rest-frame of the main emission lines on the spectra.}
    \label{fig:hb_cnt}
\end{figure*}

\begin{figure}
    \centering
    \includegraphics[width=0.8\columnwidth]{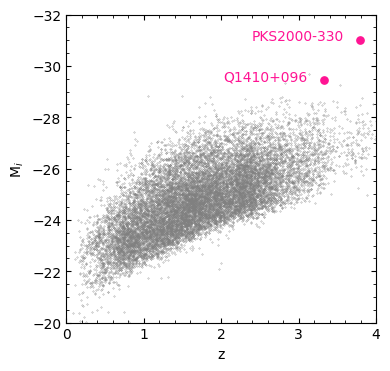}
    \caption{Location of the two sources (in pink) in the Hubble diagram. Grey dots represent a random subsample of the SDSS DR-16 catalogue from \cite{Lyke_2020}. Galactic extinctions are obtained from \cite{Schlafly_2011} and $K$-correction through \cite{Richards_2006}.}
    \label{fig:hubble_diagram}
\end{figure}

\par In this work we present measured properties in the optical and UV spectral regions of two quasars, one radio-loud (PKS200-330) and one radio-quiet (Q1410+096), selected from our ongoing work of a complete sample of 36 quasars at high redshift ($z=2-4$), that will be presented in a forthcoming paper.  The infrared observations were performed at the 8-meter VLT telescope, through the European Southern Observatory (ESO) programmes 083.B-0273(A) for PKS2000-330 and 085.B-0162(A) for the observations of Q1410+096. The ISAAC spectrograph was operated in service mode with a slit width of 0.6''. The spectroscopic reduction of the new VLT observations were performed in the standard way using the routines of the astronomical package \textsc{IRAF}. Fig. \ref{fig:hb_cnt} shows the obtained rest-frame optical spectra of the two quasars, with a S/N ratio of {$\approx$ 55} in both spectra. Redshift estimation was performed based on the H$\beta$ emission line profile fitting, and then the same $z$ was applied for both optical and UV spectra.
\par For the UV spectral range (observed in the optical domain at the redshift of the sample), which includes high ionization lines as C \textsc{IV}$\lambda$1549, He \textsc{II}$\lambda$1640, Si \textsc{IV}+O \textsc{IV}]$\lambda$1500, we use the archive from the Sloan Digital Sky Survey to obtain the spectrum of Q1410+096. In the case of PKS2000-330 (the radio-loud source) the UV spectrum was digitalized from \cite{barthel_1990}. Fluxes at radio frequencies and the study of radio data of the sources were collected from the Faint Images of the Radio Sky at Twenty-Centimeters (FIRST) survey archive, as well as from the NVSS survey. 

\par The selected sources are among the objects with the highest redshift (z) in our sample. The motivation behind selecting these two QSOs is that they share a similar MS location and a similar optical spectrum, but are different in terms of radio emission and UV spectrum. Table \ref{tab:properties} lists the coordinates (Col. 2 and 3), estimated redshift ($z$, Col. 4), $i$-band absolute magnitude ($M_{\rm{i}}$, Col. 5), and $H$-band apparent magnitude ($m_{\rm{H}}$, Col. 6) of these two sources. In Fig. \ref{fig:hubble_diagram} is shown the location of the two sources in the $M_{\rm{i}}$ vs. $z$ plane, where we have also represented a random QSO subsample from the SDSS DR16 catalogue \citep{Lyke_2020} for a comparison with our data. The two sources present higher absolute magnitude $M_{\rm{i}}$ and larger redshift when compared with the SDSS data average. 

\begin{figure*}[htp!]
    \centering
   \includegraphics[width=0.95\linewidth]{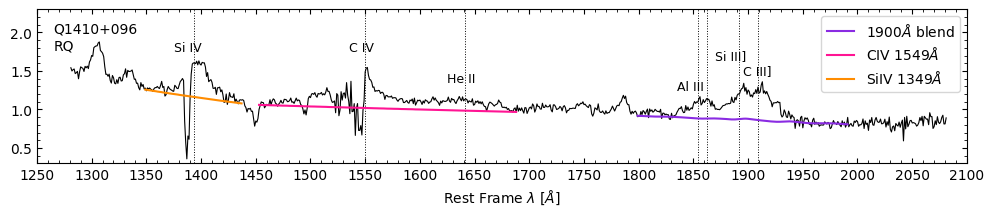}
     \\
     \includegraphics[width=0.95\linewidth]{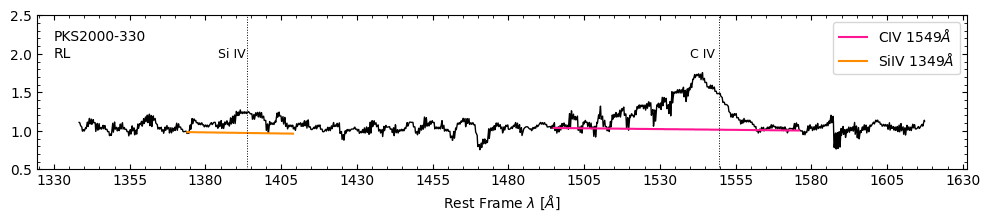}
 
    \caption{Spectra for the UV region of Q1410+096 (top panel) and PKS2000-330 (bottom panel). Orange, pink, and purple lines identify the local fitted continuum by \texttt{specfit} on the corresponding UV regions. Vertical dotted lines show the rest-frame (obtained from the H$\beta$ fittings) of the main emission lines on the spectra.}
    \label{fig:uv_cnt}
\end{figure*}

\begin{figure*}[htp!]
    \centering
    \includegraphics[width=0.3258\linewidth]{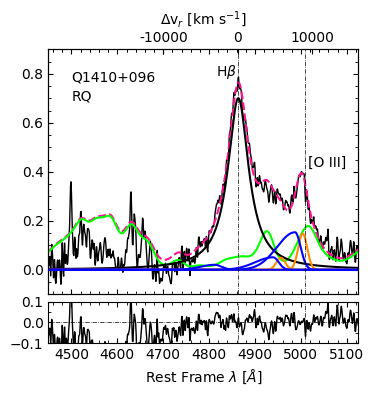}
    \includegraphics[width=0.3214\linewidth]{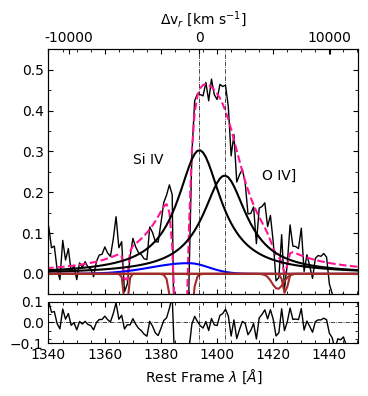}
    \includegraphics[width=0.338\linewidth]{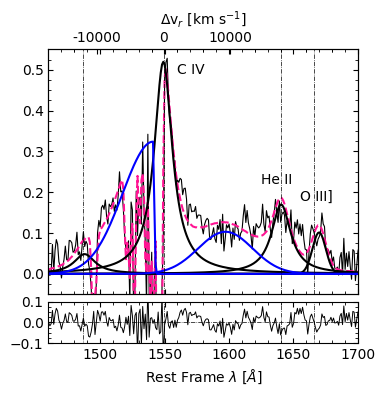}
    \\
    \includegraphics[width=0.3258\linewidth]{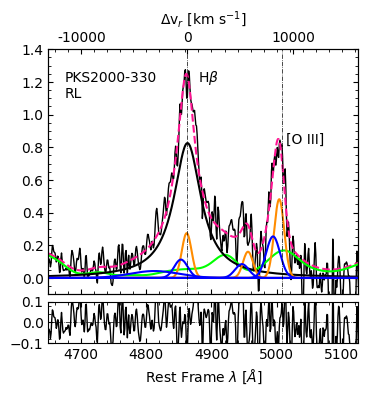}
    \includegraphics[width=0.3318\linewidth]{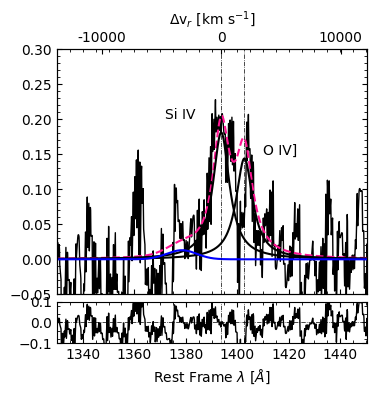}
     \includegraphics[width=0.333\linewidth]{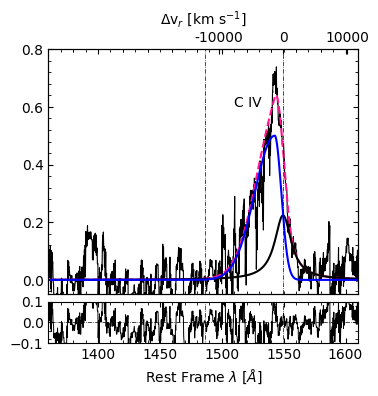}
        \caption{Fittings and residuals for the optical and UV region of Q1410+096 (top panels) and PKS2000-330 (bottom panels). \textit{From left to right:} H$\beta$+[O III]$\lambda\lambda4959,5007$, Si IV$\lambda1392$, and C IV$\lambda1549$ emission lines regions. Pink dashed lines show the final fitting. Broad components are represented by black lines, while blueshifted components are in blue. Orange and green lines represent narrow and Fe \textsc{II} components. Brown lines represent the absorptions seen in the spectra. Residuals are shown in the bottom of each plot.}
    \label{fig:uv}
\end{figure*}

\begin{itemize}

\item \textbf{Q1410+096:} this object has been identified as a Broad Absorption Line (BAL) QSO \citep{Allen_2011} with absorptions seen mainly on the region of the C\textsc{IV}$\lambda$1549 and Si\textsc{ IV}$\lambda$1397 emission lines (see Fig. \ref{fig:uv_cnt}). The UV spectrum was obtained from the SDSS archive through the DR16 catalog \citep{Ahumada_2020}. The  Q1410+096 redshift  is reported in the SDSS database as 3.3405, which is significantly different (by $\delta z \sim$ 0.0165) from our estimation in the optical spectrum, $z \approx 3.3240$. We have verified that there are no calibration problems in the optical spectrum. In addition, our determination of the redshift is based on the fitting of the H$\beta$ line, showing a peaked and well-defined profile (see Fig. \ref{fig:uv} upper left) in which the uncertainty in the centroid of the H$\beta$ is about 100 km s$^{-1}$. We have also carried out a fitting of the 1900{\r{A}} blend in the UV SDSS spectrum and we have verified that (as can be seen in Fig. \ref{fig:uv_cnt} upper, where rest-frame position of the emission lines of the 1900\r{A} blend are shown) the redshift determined through H$\beta$ is in very good agreement with the one we obtained from the main emission lines of the 1900\r{A} blend (Al III$\lambda 1857$ doublet, Si III]$\lambda 1892$ and C III]$\lambda 1908$). Probably the different redshift in the SDSS database could be due to some error in the redshift estimation of the SDSS DR8 automatic pipeline for this object.

\item \textbf{PKS2000-330:} This source is a very powerful radio-loud quasar, presenting a radio flux of $446.0\pm15.7$ mJy according to the NVSS catalog \citep{Condon_1998}, and a corresponding radio-loudness parameter $R_{\rm{k}} = 4118$, computed following the approach introduced by \cite{Kellermann_1989}. No emission information is provided by the FIRST catalog as the position of this source is not in the sky area covered by the FIRST survey. There is also no available spectrum for this object in the SDSS database. For doing the UV spectral analysis, we digitalized the one provided by \cite{barthel_1990}. 
This spectrum has a smaller wavelength range than the one of Q1410+096, and it does not cover 1900\r{A} blend nor He\textsc{II}$\lambda$1640 emission line. We therefore analysed only Si\textsc{IV}$\lambda$1397 and C \textsc{IV}$\lambda$1549 regions. Regarding the optical region, PKS2000-330 present a very peaked and Lorentzian-like H$\beta$ profile, which also allows for a good estimation of the rest-frame. 

\end{itemize}

\section{Spectral Analysis}\label{sec3}


\par  The spectral analysis was carried out through the \textsc{specfit} routine of \textsc{IRAF}. By using this tool, we were able to perform simultaneously a minimum-$\chi^2$ fit of the continuum, the Fe\textsc{II} pseudo-continuum and the emission/absorption individual spectral line components.  In the regions of interest, we have fit for the two sources the H$\beta$+[O\textsc{III}]$\lambda\lambda$4959,5007{\r{A}} lines in the optical range, and SiIV$\lambda$1397+OIV$\lambda$1402 and   CIV$\lambda$1549+HeII$\lambda$1640 emissions (only CIV in the case of PKS2000-330 as its spectrum does not cover the HeII line) in the UV spectra.

\subsection{\textbf{Optical range}}
In the \textsc{specfit} fits of the optical region we adopted a power law to describe the continuum. For the Fe II emission, located at both sides of H$\beta$,  we used a scalable Fe\textsc{II} template as is explained in \cite{Marziani_2009}. Since Q1410+096 and PKS2000-330 are Population A quasars, the full broad profile of H$\beta$ of the two objects has been modeled including initially three components (see e.g. \cite{Marziani_2018}): 1) a symmetric and unshifted Lorentzian-like profile broad component (BC, in order to account for the virialized subsystem of the BLR); 2) a blueshifted gaussian component (BLUE), sometimes skewed, as is the case of Q1410+096. This component is mainly detected in Pop. A quasars and it is  related with non-virial motions and/or outflowing gas. Usually, when its presence is needed, it is fitted with the same FWHM and shift of the ones observed in the [OIII] lines; and 3) a narrower gaussian component (NC) consistently modeled  with the narrow  component of [OIII].

\par In the case of the  [O\textsc{III}]$\lambda\lambda$4959,5007 emission lines, each line of the doublet is fitted by a "narrow" symmetrical gaussian component (NC), and an added semibroad blueshifted gaussian component (named BLUE) associated to the presence of outflows. We fitted the doublet considering a fixed ratio of theoretical intensities of 1:3 \citep{dimitrijevic_2007}, the same FWHM, and same line shift. As in H$\beta$, the [O \textsc{III}]$\lambda\lambda$4959,5007 BLUE component are allowed to be skewed to the blue, indicating the presence of an outflow contribution. Also, our initial guess is that the blueshifted components of H$\beta$ and [O \textsc{III}]$\lambda\lambda$4959,5007 share the same shift, FWHM and asymmetry.

\begin{center}
\begin{table}[t]%
\centering
\caption{Flux intensities of the H$\beta$ and C \textsc{IV}$\lambda$1549 emission lines.\label{tab_flux}}%
\tabcolsep=7pt%
\begin{small}
\resizebox{\linewidth}{!}{
\begin{tabular*}{15.5pc}{lcr}
\toprule
\hline
\textbf{Source} & \textbf{f(H$\beta$)} & \textbf{f(C \textsc{IV}$\lambda$1549})\\
 & [10$^{-15}$ erg s$^{-1}$ cm$^{-2}$ \r{A}$^{-1}$] & [10$^{-15}$ erg s$^{-1}$ cm$^{-2}$ \r{A}$^{-1}$]\\
(1) & (2) & (3) \\
\midrule
\textbf{Q1410+096} & 5.9 & 2.2\\
\textbf{PKS2000-330} & 6.6 & 4.5\\

\hline

\bottomrule
\label{tab:measurements}
\end{tabular*}
}
\end{small}
\end{table}
\end{center}

\subsection{\textbf{UV range}}
Regarding the fitting of the UV spectra, we divide the spectral range into two regions centered on the most important emission lines: one in which C\textsc{IV}$\lambda$1549 dominates the emission,  and the other centered in Si\textsc{IV}$\lambda$1397+O\textsc{IV}$\lambda$1402  lines. In both cases the continuum was modeled locally by a power-law as it is shown in Fig. \ref{fig:uv_cnt}. Similarly to the H$\beta$ line, broad UV lines are fitted by using an unshifted Lorentzian profile (BC), kept fixed at the rest-frame determined with H$\beta$, plus one or more asymmetric Gaussians to model the blueward excess (BLUE).
\par In the case of Q1410+096 where both CIV and SiIV are contaminated  by the  presence of strong absorption features in the blue side, the fittings include  absorption lines modeled as Gaussians, so that \textsc{specfit} takes into account them to evaluate the broad emission lines parameters. Nevertheless, the contamination by strong absorption features in the blue side of both profiles make the fit less accurate and the fitted parameters affected by larger uncertainties. Since C IV$\lambda$1549 and He II$\lambda$1640 are expected to present a similar shape, the spectral analysis of the He II$\lambda$1640 emission line in Q1410+096 is performed in a similar way of C IV$\lambda$1549, sharing the same FWHM and shifts \citep{martinez_aldama_2018}. In the same way, broad component of SiIV was modeled with the same emission components of CIV (BC+BLUE), with only the flux varying freely.


\section{Results}\label{sec4}
\subsection{H$\beta$+[O \textsc{III}]$\lambda\lambda4959,5507$}

\par Fig. \ref{fig:hb_cnt} shows the VLT spectra used on the H$\beta$+[O \textsc{III}]$\lambda\lambda$4959,5007. 
The two objects show very similar behaviour on the optical range, specially in terms of Fe \textsc{II} contribution (green lines) and H$\beta$ profile. Flux intensities of H$\beta$ are shown in Table \ref{tab_flux}. \textsc{specfit} models for the optical data of the two sources are shown in the left plots of Fig. \ref{fig:uv}. The FWHM of the H$\beta$ broad component of Q1410+096 and PKS2000-330 are also very similar, however the shape of the profile at the rest-frame wavelength present a small difference making it necessary to include a very narrow component to account for the narrowed peak seen in the profile of PKS2000-330.

\par Table \ref{tab:measurements} lists the measurements we obtained after the \textsc{specfit} fitting for the optical region. We present FWHM and centroid velocity at 1/2 intensity (c(1/2)) of the full profile for H$\beta$ and [O \textsc{III}]$\lambda\lambda$4959,5007 as well as the relative intensity compared to the total flux (I/I$_{\rm{tot}}$), and the FWHM and wavelength shift for the individual line components. There are no shifts for the broad line components of H$\beta$. Also, we do not include the blueshifted components of the H$\beta$ profile on the full profile measures because they are too small to present a significant contribution to the full emission line or are clearly related with the [O \textsc{III}]$\lambda\lambda$4959,5007 blueshifted components (case in which it is possible that the components present the same physical origin). In the case of the [O \textsc{III}]$\lambda\lambda$4959,5007 emission lines, the blueshifted components are stronger and present a more skewed shape for Q1410+096 than for PKS2000-330.

\par
\begin{figure}
    \centering
    \includegraphics[width=\columnwidth]{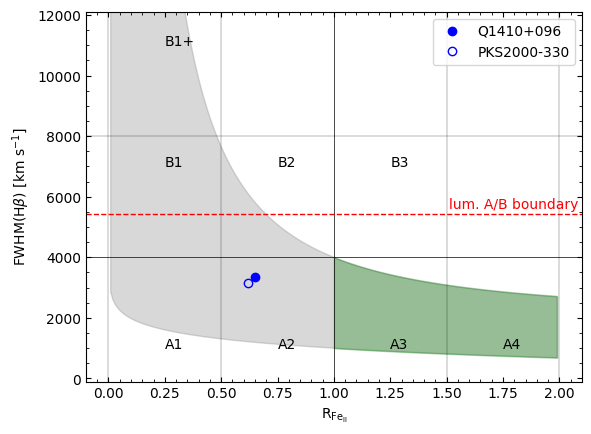}
    \caption{Location of the two sources on the optical plane of the 4DE1. The radio-quiet quasar (Q1410+096) is represented by the filled circle, while the radio-loud quasar (PKS2000-330) is shown as an open circle. Grey and green regions indicate the distribution of the Main Sequence of quasars. Red horizontal line shows the Population A/B boundary for high luminosity quasars, following the FWHM(H$\beta$)-M$_{\rm{i}}$ relation from \cite{Sulentic_2004}.}
    \label{fig:optical_plane}
\end{figure}

 After performing the multicomponent fitting, we can locate the two sources on the optical plane of the MS. Fig. \ref{fig:optical_plane} shows the location of Q1410+096 and PKS2000-300 on the Main Sequence (represented by the grey and green areas). The red dashed line indicates the A/B Population boundary for the FWHM in the Main Sequence corresponding to a quasar luminosity of about $10^{48}$ erg s$^{-1}$, representative of the luminosity of our sample,  according to \cite{Marziani_2009}. As can be seen, the two sources are very similar in terms of FWHM(H$\beta$) and R$_{\rm{Fe \textsc{II}}}$ and are classified as Pop. A2 quasars.

\subsection{UV}

 UV spectra and fittings of the C\textsc{IV}$\lambda$1549+He \textsc{II}$\lambda$1640 and SiIV$\lambda$1392 are shown in Figs. \ref{fig:uv_cnt} and  \ref{fig:uv}, respectively. Table \ref{tab:measurements_uv} reports the FWHM and c(1/2) of the CIV full profile as well as the intensities, FWHM, and shift of the individual components. The blueshifted component of the C\textsc{IV}$\lambda$1549 is usually much stronger than the ones observed in the optical range, which can be seen as an indicative of wind activities surrounding the central region. This is also observed in the two sources, where the full profile of CIV is blueshifted by about 2000 km s$^{-1}$. In the case of the RQ BAL Q1410+096 the presence of strong absorption features could explain the lower blue-shift observed (parametrized by c(1/2)) in the CIV full profile when compared with the RL PKS2000-330. Previous studies had shown that smaller values of c(1/2) in UV lines in BALs are due mainly to the presence of deep absorptions in the blue side of the profiles (\cite{martinez_aldama_2018}). 

\par Also, when comparing the two sources in this region, it is clear that their spectral behaviour is different. While in Q1410+096 the ratio between the intensities of C\textsc{IV}$\lambda$1594 and Si\textsc{IV}$\lambda$1397 is $\sim$ 0.96 (a value expected for Pop. A quasars), the situation is the opposite for PKS2000-330. In this case, there is a very small contribution of Si\textsc{IV}$\lambda$1397, much smaller than the C\textsc{IV}$\lambda$1549 profile, presenting a Pop. B-like UV spectrum with a C\textsc{IV}$\lambda$1594/Si\textsc{IV}$\lambda$1397 $\sim$ 1.40.

\begin{center}
\begin{table*}%
\centering
\caption{Measurements for the optical range.\label{tab2}}%
\tabcolsep=2.25pt%
\begin{normalsize}
\resizebox{\linewidth}{!}{
\begin{tabular*}{49.5pc}{lccccccccccccccccccccr}
\toprule
\hline

& & &\multicolumn{17}{c}{Optical}\\
\cline{2-21}
Source & &\multicolumn{2}{c}{H$\beta_{\rm{full}}$} & & \multicolumn{2}{c}{H$\beta_{\rm{BC}}$} & & \multicolumn{2}{c}{H$\beta_{\rm{NC}}$} & &\multicolumn{2}{c}{[O \textsc{III}]$_{\rm{full}}$}  & & \multicolumn{3}{c}{[O \textsc{III}]$_{\rm{BLUE}}$} & & \multicolumn{3}{c}{[O \textsc{III}]$_{\rm{NC}}$}\\
\cline{2-4} \cline{6-7} \cline{9-10} \cline{12-13} \cline{15-17} \cline{19-21} 
& & \textbf{FWHM} & \textbf{c(1/2)} & & \textbf{I/I$_{\rm{tot}}$} & \textbf{FWHM} &  & \textbf{I/I$_{\rm{tot}}$} & \textbf{FWHM} & & \textbf{FWHM} & \textbf{c(1/2)} & & \textbf{I/I$_{\rm{tot}}$} & \textbf{FWHM} & \textbf{Shift} & & \textbf{I/I$_{\rm{tot}}$} & \textbf{FWHM} & \textbf{Shift}\\
& & \textbf{[km s$^{-1}$]} & \textbf{[km s$^{-1}$]} & & & \textbf{[km s$^{-1}$]} & & & \textbf{[km s$^{-1}$]} & & \textbf{[km s$^{-1}$]} & \textbf{[km s$^{-1}$]} & & & \textbf{[km s$^{-1}$]} & \textbf{[km s$^{-1}$]} & & & \textbf{[km s$^{-1}$]} & \textbf{[km s$^{-1}$]}\\
(1) & & (2) & (3) & & (4) & (5) & & (6) & (7) & & (8) & (9) & & (10) & (11) & (12) & & (13) & (14) & (15)\\
\midrule
\textbf{Q1410+096} & & 3394 & 54 & & 1.0 & 3394 & & - & - & & 3363 & -1404 & & 0.73 & 5573 & -1255 & & 0.27 & 1379 & -260 \\
\textbf{PKS2000-330} & & 3138 & 28 & & 0.93 & 3138 & & 0.072 & 1082 & & 1314 & -425 & & 0.42 & 1550 & -827 & & 0.58 & 1082 & -263 \\

\hline

\bottomrule
\label{tab:measurements}
\end{tabular*}
}
\end{normalsize}
\end{table*}
\end{center}

\begin{center}
\begin{table}%
\centering
\caption{Measurements for the C IV$\lambda$1549.\label{tab2}}%
\tabcolsep=2.25pt%
\begin{normalsize}
\resizebox{\linewidth}{!}{
\begin{tabular*}{27pc}{lccccccccr}
\toprule
\hline

 & \multicolumn{9}{c}{UV}\\
\cline{2-10} 
Source & \multicolumn{2}{c}{C \textsc{IV}$_{\rm{full}}$} & & \multicolumn{3}{c}{C \textsc{IV}$_{\rm{BLUE}}$} & & \multicolumn{2}{c}{C \textsc{IV}$_{\rm{BC}}$}\\
\cline{2-3} \cline{5-7} \cline{9-10}
 & \textbf{FWHM} & \textbf{c(1/2)} & & \textbf{I/I$_{\rm{tot}}$} & \textbf{FWHM} & \textbf{Shift} & & \textbf{I/I$_{\rm{tot}}$} & \textbf{FWHM} \\
 & \textbf{[km s$^{-1}$]} & \textbf{[km s$^{-1}$]} & & & \textbf{[km s$^{-1}$]} & \textbf{[km s$^{-1}$]} & & & \textbf{[km s$^{-1}$]}\\
(1) & (2) & (3) & & (4) & (5)  & (6) & & (7) & (8) \\
\midrule
\textbf{Q1410+096} & 6311 & -1746 & & 0.42 & 10882 & -1525 & & 0.58 & 3293\\
\textbf{PKS2000-330} & 4950 & -1923 & & 0.71 & 7339 & -1219 & & 0.29 & 3141\\

\hline

\bottomrule
\label{tab:measurements_uv}
\end{tabular*}
}
\end{normalsize}
\end{table}
\end{center}

\section{Discussion: Is there any relation between H$\beta$+[O \textsc{III}]$\lambda\lambda$4959,5007 and C IV$\lambda1549$?}
\par By analysing Table \ref{tab:measurements_uv}, it is clear that the blueshift component plays an important role in the full profile of C \textsc{IV}$\lambda1549$ in the two sources, with a comparable shift towards lower wavelengths ($\sim$ 2000 km s$^{-1}$). Q1410+096 (radio-quiet) shows a C \textsc{IV}$\lambda1549$ BLUE with a FWHM $=10882$ km s$^{-1}$, while PKS2000-330 (radio-loud) have a C \textsc{IV}$_{\rm{BLUE}}$ that is somewhat narrower ($\sim 7300$ km s$^{-1}$). The fact that radio-quiet sources show a preference of having wider FWHM(C \textsc{IV}$_{\rm{BLUE}}$) than the radio-loud ones is already known by the recent literature and seems to be the normal behaviour in this type of sources \citep{Richards_2021}. Apart from that, these results also may indicate that there is an important contribution of outflowing gas in the nuclear activity in the two sources, in agreement with previous results in other quasar samples \citep{Marziani_2017, sulentic_2017}.

\par Another point that we would like to highlight here is that the radio-quiet we analyse presents a slightly higher shift to blue wavelengths when compared to the radio-loud. However, this behaviour is not exclusively observed in C \textsc{IV}$\lambda$1549. Actually, it is also seen in the blueshifted component of [O \textsc{III}]$\lambda\lambda$4959,5007 \citep{Ganci_2019}, although rare radio-loud NLSy1s do show significant shifts in [O \textsc{III}] \citep{bertonetal16}. In the two sources, C \textsc{IV}$\lambda$1549 and [O \textsc{III}]$\lambda\lambda$4959,5007 present similar shifts towards the blue. As a consequence, the full profiles of these two emission lines present a negative c(1/2), indicating a strong contribution of the blueshifted components to the full profile. Also, the FWHM(C\textsc{IV}$_{\rm{full}}$) is much wider for the radio-quiet source than for the radio-loud source, with a difference greater than 1000 km s$^{-1}$ for C \textsc{IV}$\lambda$1549 and $\sim 2000$ km s$^{-1}$ for [O \textsc{III}]$\lambda\lambda$4959,5007.

\par Both [O \textsc{III}]$\lambda\lambda$4959,5007 and C \textsc{IV}$\lambda$1549 blueshifted components represent a significant fraction of their respective full profiles. This is an expected behaviour especially at high/intermediate redshift \citep{Sulentic_2004, Sulentic_2007, Marziani_2009}. When compared to samples with lower redshift,  sources such as Q1410+096 and PKS2000-330 (and in general sources at intermediate redshifts) tend to show  components in lines like [O \textsc{III}]$\lambda\lambda$4959,5007 and C \textsc{IV}$\lambda$1549 that are wider and more blueshifted.

\par \cite{sulentic_2017}, for instance, analyse samples with different redshifts and luminosities showing that the higher the luminosity, the stronger the CIV outflows tend to be. In the high-$z$ context, the emission component related with the outflowing gas becomes the dominant constituent of the broad profile. Similar behaviour is observed in [O III] emission lines, which present significant differences in the full profile when comparing high and low luminosity sources. \cite{canodiaz_2012} find a massive outflow on kpc scales in the [O III] for 2QZJ002830.4-281706, which has a $z=2.4$. Similar results were also found in \cite{marziani_2016}, for a Hamburg ESO sample with intermediate $z$ and high luminosity. \cite{zakamska_2016} analyse four extremely red quasars with $z \sim 2.5$ and discover blueshifted components with very broad FWHM and strongly shifted towards blue wavelengths.


\par The FWHM of the broad components of H$\beta$ and C \textsc{IV}$\lambda$1549 also indicate some correspondence between optical and UV regions. For Q1410+096, the FWHM of the two lines differ by $\sim 100$ km s$^{-1}$. And in the case of PKS2000-330, this difference is smaller than 5 km s$^{-1}$. When considering the full profile, it can be well reproduced by almost only the BC component in the H$\beta$ profile. This is not true for C \textsc{IV}$\lambda$1549, case in which the FWHM of the broad component represent only 58\% for Q1410+096 and 29\% for PKS2000-330.





\section{Conclusions}\label{sec5}
\par We analysed a pair of Pop. A2 quasars, which are similar respect to the optical spectra and are very close on the MS classification but present significant differences on the UV data and also on the radio emission: one is a very powerful radio-loud QSO (PKS2000-330) and the other is a radio quiet (Q1410+096). The main conclusions are:
\begin{itemize}
    \item The C \textsc{IV}$\lambda$1549 and [O \textsc{III}]$\lambda\lambda$4959,5007 emission line profiles present a similar behaviour in Q1410+096 and also in PKS2000-330, with a more significant shift towards lower wavelengths in the case of the radio-quiet quasar;
    
    \item The [O \textsc{III}]$\lambda\lambda$4959,5007 emission lines in the two sources are different. In PKS2000-330, we observe a symmetric blueshifted component. On the other hand, in Q1410+096 it is necessary a very skewed blueshifted component (also in H$\beta$). This may be an indicative that the two objects probably have different behaviour in the observed winds; 
    
    \item The general behaviour of the UV spectra is very different when comparing the two sources. In the case of the radio-quiet source (Q1410+096) C \textsc{IV}$\lambda$1549 and Si \textsc{IV}$\lambda1392$ are very similar in intensity and shape, which is a characteristic expected for Pop. A quasars. The same is not true for PKS2000-330, which is a radio-loud quasar and present a more Pop. B-like spectrum in the UV region.
    
\end{itemize}

However, a more detailed study is needed in order to perform a complete analysis of the clear behaviour of radio-quiet and radio-loud sources along the Main Sequence. The next step of this work will be to perform an analysis of the spectroscopic properties of our complete sample of 36 quasars and to study the differences between radio-loud and radio-quiet sources.

\section*{Acknowledgments}
The authors thank the anonymous referee for her/his valuable suggestions that helped us to significantly improve the present paper. A.D.M.  acknowledges  the  support  of  the  INPhINIT  fellowship  from  "la  Caixa"  Foundation  (ID 100010434).  The  fellowship  code  is  LCF/BQ/DI19/11730018.  A.D.M.  and  A.d.O.  acknowledge financial  support  from  the State Agency for Research of the Spanish MCIU through the project PID2019-106027GB-C41 and the “Center of Excellence Severo Ochoa” award to the Instituto de Astrofísica de Andalucía (SEV-2017-0709).











\nocite{*}
\bibliography{Wiley-ASNA}%

\section*{Author Biography}

\begin{biography}{
\includegraphics{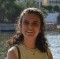}}
{\textbf{Alice Deconto Machado.} PhD candidate at the Instituto de Astrofísica de Andalucía (IAA-CSIC, Granada, Spain), under the supervision of Dr. Ascensión del Olmo Orozco (IAA-CSIC, Granada, Spain) and Dr. Paola Marziani (INAF-Padova, Italy).}
\end{biography}

\end{document}